\documentclass[twocolumn,superscriptaddress,pre,showpacs]{revtex4-1}

\usepackage[colorlinks,citecolor=blue,linkcolor=blue]{hyperref}

\usepackage[T1]{fontenc}
\usepackage{times}

\usepackage{amsmath}
\usepackage{amsfonts}
\usepackage{amssymb}

\usepackage{microtype}

\usepackage{graphicx}

\newcommand{\rmd}{\mathrm{d}}  

\newcounter{myParagraph}
\renewcommand{\paragraph}[1]{\stepcounter{myParagraph}{\it(\alph{myParagraph})} \emph{#1}}

\begin{document}

\title{Soft bounds on diffusion produce skewed distributions and Gompertz growth}

\author{Salvatore Mandr\`a}
\affiliation{Department of Chemistry and Chemical Biology, Harvard University, 12 Oxford Street, Cambridge MA, USA}
\affiliation{Dipartimento di Fisica, Universit\`a degli Studi di Milano, via Celoria 16, 20133 Milano, Italy}
\affiliation{Istituto Nazionale di Fisica Nucleare, sezione di Milano, via Celoria 16, 20133 Milano, Italy}

\author{Marco Cosentino Lagomarsino}
\affiliation{Sorbonne Universit\'es, UPMC Univ Paris 06, UMR 7238, Computational and Quantitative Biology,\\
 15 rue de l'\'{E}cole de M\'{e}decine Paris, France}
\affiliation{CNRS, UMR 7238, Paris, France}

\author{Marco Gherardi}
\affiliation{Dipartimento di Fisica, Universit\`a degli Studi di Milano, via Celoria 16, 20133 Milano, Italy}
\affiliation{Istituto Nazionale di Fisica Nucleare, sezione di Milano, via Celoria 16, 20133 Milano, Italy}

\begin{abstract}
  Constraints can affect dramatically the behavior of diffusion
  processes.  Recently, we analyzed a natural and a technological
  system and reported that they perform diffusion-like discrete steps
  displaying a peculiar constraint, whereby the increments of the
  diffusing variable are subject to configuration-dependent bounds.
  This work explores theoretically some of the revealing landmarks of
  such phenomenology, termed ``soft bound''.
  At long times, the system reaches a steady state irreversibly (i.e.,
  violating detailed balance), characterized by a skewed ``shoulder''
  in the density distribution, and by a net local probability flux,
  which has entropic origin.  The largest point in the support of the
  distribution follows a saturating dynamics, expressed by the
  Gompertz law, in line with empirical observations.  Finally, we
  propose a generic allometric scaling for the origin of soft
  bounds. These findings shed light on the impact on a system of such
  ``scaling'' constraint and on its possible generating mechanisms.
\end{abstract}


\maketitle

\section{Introduction}
\vspace{-0.3cm}

Random processes describe a wide spectrum of phenomena in complex
systems \cite{Simon:1962}.  Diffusion processes, for instance, are used
to understand trajectories of one- or multi-dimensional fluctuating
observables or order parameters in a great variety of contexts within
and outside physics.  The validity of such diffusive descriptions --
often applicable with impressive precision to real-world systems -- is
based on the assumption that the probability of future events depends
only on the present state of the system~\cite{Risken:BOOK}.
Constraints have an influence on such processes, as they limit the
phase space that can be reached from a given state.  

In one-dimensional diffusion processes, constraints on diffusing
quantities are typically embodied by \emph{hard
  bounds}, i.e., by strict limits that the diffusing variable cannot
overcome, irrespectively of how close or distant to the boundaries the
variable already is.
The presence of bounds can alter qualitatively the properties of a
system. It is well known that absorbing or reflecting boundary
conditions affect the basic properties or a random walk, e.g. creating
steady-state distributions and affecting first-passage
times. Additional more subtle and intriguing phenomena may emerge with
hard bounds.  For example, a random (diffusional) one-dimensional
multiplicative process, in presence of a lower bound, can give rise to
power-law distributions for the value of the variable at a given
time~\cite{SornetteCont:1997,SolomonLevy:1996,AitchisonBrown:BOOK}.  

However, one can imagine a system where the constraints are not
embodied by hard bounds, but follow a different type of behavior. For
example discrete diffusion steps may be limited differently depending
on the configuration they start from.  Recently, we found empirical
evidence of precisely this behavior, which we termed \emph{soft
  bound}~\cite{Gherardi:2013PNAS}.
Note that while a classical physical example of constrained diffusion
is a Brownian particle confined in a box, whose motion is
  continuous in time, in other (often less tangible) examples, such
as stock prices~\cite{MantegnaStanley:BOOK} or the population of a
city~\cite{Batty:BOOK}, the quantity of interest is naturally measured
at discrete time intervals, and its evolution is best described by
finite-sized discrete jumps.
As we will see, the difference between hard and soft bounds is
relevant for time-discrete diffusion processes and we will thus
consider this case here.

The scope of this work is to explore some of the basic theoretical
consequences of diffusion with soft bounds.  It is important to
  stress that the precise definition of a soft bound --- as we
  formulate it here --- is motivated by compelling empirical evidence
  \cite{Gherardi:2013PNAS}.  The reasons for the emergence of such a
  behavior constitute a partially unsolved problem (see the Discussion
  section) and are not the main focus here.  Instead, the motivation
  for the present study is an exploration of its effects. 
In particular, we present a description of the general features of the
stationary state, aided by the analytical solution in a simple case,
and we show that a soft upper bound on diffusion causes a slowly
saturating dynamics for the maximum.  Specifically, we show that
detailed balance is broken, and a net local probability current of
purely entropic origin is established (which suggests a novel
rationalization of the ``Cope's law,'' a much debated feature of body
mass evolution).  Additionally, the steady-state distribution under a
soft bound cannot be obtained from a model with a hard bound and an
effective drift, and thus has to be regarded as qualitatively distinct
from previously known phenomenology. Finally, the dynamics of the
maximum follows exactly the Gompertz function, a generalized logistic
curve used in diverse contexts and observed empirically.  These
findings can be used to recognize soft bounds in real-world
systems. Finally we argue how a generic allometric scaling mechanism
can generate soft bounds.

\section{Background. Definition of soft bound and empirical evidence}
\vspace{-0.3cm}

\begin{figure}[b]
\centering
\includegraphics[scale=0.48]{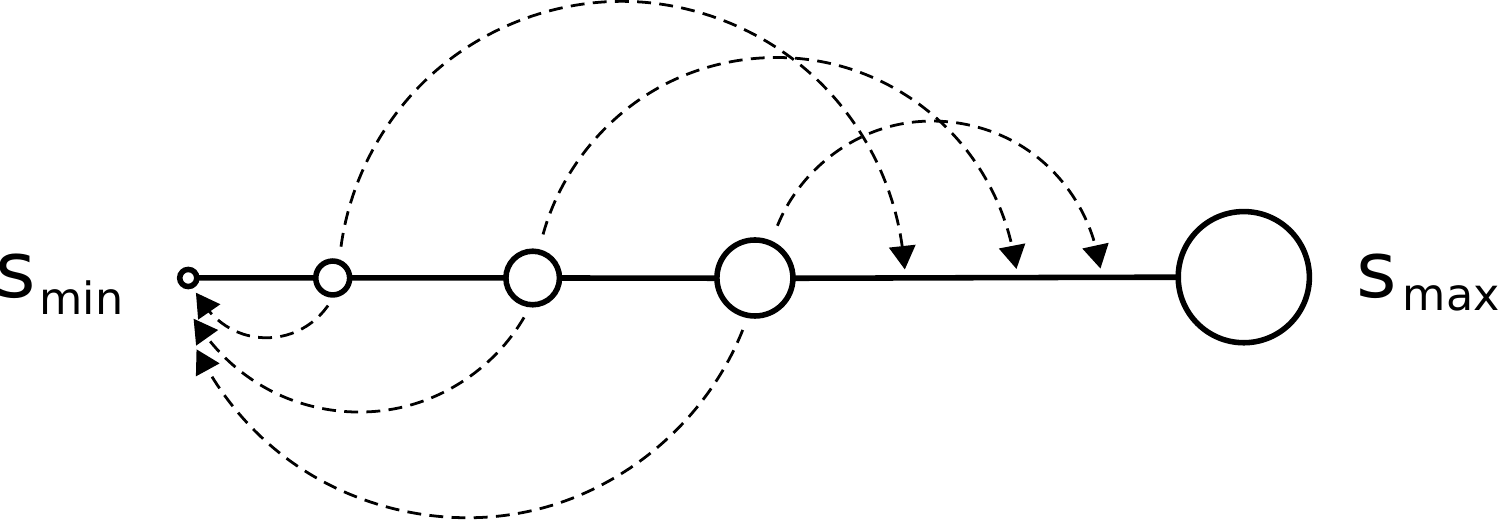}
\caption{\label{figure:softbound} Illustration of the ``soft
    bound'' mechanism. In the drawing, the dashed lines stand for the
  extremal attainable size $s$ in either direction (represented by
  both circle size and position along the interval) in a single jump,
  from different initial conditions. Jumps within these limits follow
  an assigned jump-size distribution.  The left-hand side of the
  interval is conditioned by a conventional hard bound: the minimal
  size $s_\mathrm{min}$ can be reached in a single step starting from
  any initial size.  By contrast, the right-hand part of the interval
  is limited by a soft bound, as the maximum size attainable in one
  step depends on the starting size. As a consequence, the absolute
  maximum $s_\mathrm{max}$ cannot be reached in one step from any
  initial size, but it can only be approached asymptotically.}
\end{figure}

We start by introducing the concept of soft bound through a brief
review of the recent empirical evidence suggesting its existence.
The first system where the phenomenon of a soft bound
emerges from empirical data is the dynamics of software packages, the
individual bundled pieces of software that make up an operating
system, such as Linux~\cite{Gherardi:2013PNAS}.  Packages change their
sizes during their lifespan, as a result of development, maintenance,
or repackaging.  Thus, the scalar variable ``size'' recapitulates the
result of a possibly long and complex series of operations and
processes acting on a package.  A natural time interval is set by the
distance between consecutive package releases: the size $s$ (for
instance in bytes) of a given package in a release can be compared to the
size $s'$ of that same package in the following release, defining a
\emph{jump} $\Delta=s'/s$. For Ubuntu Linux packages, these jumps are
distributed in a strikingly regular way.  While the bulk of their
distribution does not depend on time nor on the starting size $s$, its
tails are cut-off in a size-dependent way. For jumps towards lower
sizes, one expects that the size of a package cannot become smaller
than some system-wide minimum $s_\mathrm{min}$, i.e. the lower
bound is a hard bound.  This
implies that $\Delta\geq s_\mathrm{min}/s$ (the lower bound on the
jumps is inversely proportional to the starting size) and this is
indeed found in data.  However, the same behavior has not been observed
for the upper tail of the distribution.  In fact, the cut-off for jumps to
larger sizes is defined by $\Delta\leq
\left(s_\mathrm{max}/s\right)^\gamma$, with  
exponent $\gamma\approx 0.5$.  This means that the larger a package
is, the larger it can become in one step, i.e., the maximum attainable
size in one step moves further away for increasingly larger packages
(Fig.~\ref{figure:softbound}).

More formally, in a one-dimensional multiplicative discrete diffusion
process limited by a lower hard bound and an upper soft bound, as
motivated by the case of Linux package size, the bounds can be
expressed by the formula
\begin{equation}
\label{eq:multiplicative_softbound}
\frac{s_\mathrm{min}}{s} \leq \frac{s'}{s} \leq
\left(\frac{s_\mathrm{max}}{s}\right)^\gamma,
\quad \mathrm{with} \; 0<\gamma<1.
\end{equation}
Regardless of the probability distribution for the jumps
$s'/s$, such a multiplicative process can be written in an additive
form by a logarithmic transformation.  By setting $y=\log s$, $x=\log
s'$, $\Lambda_\mathrm{min}=\log s_\mathrm{min}$, and
$\Lambda_\mathrm{max}=\log s_\mathrm{max}$, the hard and soft bounds
(\ref{eq:multiplicative_softbound}) are then expressed by
\begin{equation}
\label{eq:additive_softbound}
\Lambda_\mathrm{min} \leq x \leq y + \gamma\left( \Lambda_\mathrm{max} - y\right),
\end{equation}
which makes evident the fact that the lower bound
$\Lambda_\mathrm{min}$ is a hard bound while the upper one depends on the
starting point $y$.  In the following, unless otherwise specified, we
will refer to this additive version of the process.

A second empirical system that is consistent with the diffusion
under soft bounds (although data are much sparser) is the evolution of
body masses for mammalian species~\cite{Gherardi:2013PNAS}.  Here, the
time steps are fixed by the speciation events; the ``jumps'' are
realized during cladogenesis between the mass $s$ (in kilograms) of a
mother species, and the mass $s'$ of the daughter species
\cite{Clauset2008,ClausetRedner:2009}.
A notable
consequence of the assumption of a soft upper bound in this context is
that evolution of mammalian species requires longer time (i.e., more
speciation events) to attain large increases in body mass, than it
does for large decreases.  This macro-evolutionary asymmetry has been
observed in fossil data: the tendency to extreme dwarfism, for instance on
islands, is more common than the opposite
trend~\cite{EvansJones:2012}.

\section{Discrete diffusion between a hard and a soft bound}
\vspace{-0.3cm}

We now define more technically the diffusion process of interest.  The
formal framework is that of Markov chains.  To fix the notation, let
$\mathcal{P}_{y\to x}$ denote the transition kernel, i.e., the
probability to jump from position $y$ to $x$, and let $\rho_t(x)$ be
the state of the system, i.e., the density distribution of the
diffusing particles at time $t$,
e.g. the (logarithmic) package sizes or species masses.
This is an inherently discrete process, hence $t$ takes
only integer values. State space, instead, is continuous in general.
The evolution is then given by
\begin{equation}\label{eq:def1}
	\rho_{t+1}(x) = \int_{\mathbb{R}}\rmd y\,\rho_t(y)\,\mathcal{P}_{y\to x}.
\end{equation}
We will consider jump probabilities $\mathcal{P}_{y\to x}$ 
which are the superposition of two components: 
(i) an underlying transition probability $\pi(x-y)$,
which is translationally invariant and does not necessarily have a
bounded domain, and (ii) the bounding kernel $\beta(x,y)$, which can
be written in terms of a characteristic function as
\begin{equation}
\label{eq:bounding_kernel}
\beta(x,y)=\chi_{[\Lambda_\mathrm{min},y+\gamma\left(\Lambda_\mathrm{max}-y\right)]}(x).
\end{equation}
$\mathcal{P}_{y\to x}$ is then obtained by normalizing the product of
the two kernels:
\begin{equation}
\label{eq:kernel_product}
\mathcal{P}_{y\to x} = \frac{1}{\mathcal Z(y)}\pi(x-y)\beta(x,y),
\end{equation}
where $\mathcal Z(y)$ is the position-dependent normalization
\begin{equation}
\label{eq:normalization}
\mathcal Z(y) = \int_{\Lambda_\mathrm{min}}^{y+\gamma\left(\Lambda_\mathrm{max}-y\right)}
\!\!\!\pi(x-y) \,\rmd x.
\end{equation}   
The dependence of $\mathcal Z$ on $y$ makes it difficult in general to
find analytic solutions to the evolution, even in the long-time limit.

\section{Numerical and analytical characterization of the stationary state }
\vspace{-0.3cm}
\label{section:flat_distribution}

The dynamics defined above has a stationary state, i.e., a
distribution $\rho(x)$ that satisfies the following equation:
\begin{equation}\label{eq:def2}
  \rho(x) = \int_{\mathbb{R}}\rmd y\,\rho(y)\,\mathcal{P}_{y\to x}.
\end{equation}
The existence of a solution $\rho(x)$ for the above equation 
will be proven later in this section for a flat underlying 
transition probability $\pi(x-y)$. 
Numerical evidence shows that a stationary state is
reached also for more general forms of $\pi(x-y)$.

Note that the Markov chain given by the jump distribution
$\mathcal{P}_{x\to y}$ in Eq.~(\ref{eq:kernel_product}) is
irreversible, i.e., the probability of a history
$\left(x_1,\,x_2,\,\ldots,\,x_{m-1}\right)$ in general differs from
that of $\left(x_{m-1},\,x_{m-2},\,\ldots,\,x_1\right)$
\cite{Kelly:BOOK}.  This is witnessed by the violation, at the
stationary state $\rho(x)$, of the detailed balance condition,
\begin{equation}
\label{eq:detailed_balance}
\mathcal{P}_{x\to y} \,\rho(x) - \mathcal{P}_{y\to x} \,\rho(y)\equiv 0,
\end{equation}
which imposes a vanishing
probability flow between any two states $x$ and $y$.
Indeed, let $x$ and $y$ be two states 
lying in the interval $[\Lambda_{min},\,\Lambda_{max})$ and
such that
$x>y+\gamma\left(\Lambda_\mathrm{max}-y\right)$. Since
the condition in Eq.~(\ref{eq:additive_softbound}) is violated,
the bounding kernel $\beta(x,y)$ vanishes and then 
a jump from $y$ to $x$ is suppressed by Eq.~(\ref{eq:kernel_product}),
i.e. $\mathcal{P}_{y\to x}=0$.
However, $x > y$ implies $\mathcal{P}_{x\to y} \not = 0$, due to the
fact that \emph{backward} jumps are always allowed
by Eq.~(\ref{eq:additive_softbound}). Consequently,
since $\rho(x) \not = 0$ (as we will show later in this Section),
the detailed balance condition in Eq.~(\ref{eq:detailed_balance})
can never be satisfied.
A more detailed analysis based on entropy production, 
not relying on the vanishing of $\mathcal{P}$
and only involving two points for which $\beta(x,y)\neq 0$,
will be given later in this section.

We explored the properties of the stationary state by analytical and
numerical calculations as well as by computer simulations.  An
analytical approach is unfeasible in general, but it can be carried
out in the special case where the transition probability $\pi(x-y)$ is
flat between the hard lower bound and the soft upper bound.  The
salient qualitative features of the steady state realized in this
special case do not change if a Gaussian distribution is chosen for
$\pi(x-y)$ (see below).  To investigate the cases where $\pi(x-y)$ is
not flat, one has to take a numerical approach.

\paragraph{Numerical solutions.} 
We took two different numerical approaches to the solution of the
problem: direct numerical integration and Monte Carlo simulations.  A
numerical approximation of the density distribution $\rho_t(x)$ at
times $t=0,1,\ldots$ can be obtained by iterative integration,
starting from an initial density distribution $\rho_0(x)$, by means of
Eq.~(\ref{eq:def1}).  As $\mathcal{P}_{y\to x}$ is different from zero
only for
$x\in\left[\Lambda_\mathrm{min},\,\Lambda_\mathrm{max}\right]$, so
will be the density distribution $\rho_t(x)$, provided that
$\rho_0(x)$ is supported on the same interval.  We defined a spatial
discretization of $\rho_t(x)$ as follows.  Let $\delta x$ be a fixed
and small integration step, and define $x_k = \Lambda_{min} +
k\,\delta x$, with $k=0,1,\ldots,M$, where
$M=\lfloor\left(\Lambda_\mathrm{max}-\Lambda_\mathrm{min}\right)/\delta
y\rfloor$ is the largest value of $k$ for which $x_k\leq
\Lambda_\mathrm{max}$.  Then the discretized density distribution
$\rho_1(x_i)$ at time $t = 1$ can be computed from
the discretized version of Eq.~(\ref{eq:def1}), using the trapezoidal
rule.  The density distribution at successive time steps is then
obtained by iterative application of
this procedure.

The Monte Carlo method is based instead on an implementation of the
microscopic processes that lead to the continuum description in
Eq.~(\ref{eq:def1}).  In practice, we used a pool of $N$ uncorrelated
``particles,'' whose positions $y_i$ ($i$ is now the particle index)
evolve in discrete time steps by following the jump distribution
$\mathcal{P}_{y\to x}$.  In the following we will choose a
delta-shaped initial condition, which translates, at time $t = 0$, to
all particles being displaced at the same position $x(0)$.  At later
times, the new position $x_i(t+1) = x_i(t) + \Delta$ for the $i$-th
particle is randomly chosen according to the jump distribution
$\mathcal{P}$:
\begin{equation}
\label{eq:montecarlo}
\Delta \sim \mathcal{P}_{x_i(t)\to x_i(t)+\Delta}.
\end{equation}
Note that the probability density for the variable $\Delta$ depends
explicitly on the position $x_i(t)$ of the $i$-th particle at time
$t$.  If the number of particles $N$ is sufficiently large, the
density distribution $\rho_t(x)$ can be sampled by a histogram, i.e. a
count of the number of particles in the interval $\left[x,x+\delta
  x\right]$.

Although the two methods are different --- numerical integration
being deterministic, and Monte Carlo simulation
(\ref{eq:montecarlo}) being intrinsically stochastic --- they are
expected to attain the same results in their ``thermodynamic'' limits
$N\to\infty$ and $\delta y \to 0$.  However, since extracting random
numbers is computationally more demanding than computing the sum of
real numbers, numerical integration results to be faster and more
precise than the stochastic method in this situation (however, this is
not always the case when different time scales compete, see
e.g. Ref.~\cite{GherardiJourdan}).

\paragraph{Analytical solution for flat transition probability.}
Let us consider a flat transition probability
$\pi(x-y)=\mathrm{const}$.  This gives rise to the simplest possible
form of $\mathcal{P}$ with a soft bound.  Summing up the definitions
given in Eqs.~(\ref{eq:bounding_kernel})--(\ref{eq:normalization}), we obtain the following piecewise
continuous function, which for brevity we will term ``box
distribution'':
\begin{equation}\label{def:box}
	\mathcal{P}_{y\to x} = 
	\begin{cases}
	\mathcal{Z}^{-1} \quad &\Lambda_\mathrm{min} < x < y + \gamma(\Lambda_\mathrm{max} - y)\\
	0 & \text{otherwise},
	\end{cases}
\end{equation}
where $\mathcal{Z} = {y+\gamma(\Lambda_\mathrm{max}-y) -
  \Lambda_\mathrm{min}}$.  The stationary distribution $\rho(x)$
satisfies the definition, Eq.~(\ref{eq:def2}), which then assumes the
form
\begin{equation}\label{eq:def3}
	\rho(x) = \int_{y_\mathrm{inf}}^{\Lambda_\mathrm{max}} 
		\frac{\rho(y)}{y+\gamma(\Lambda_\mathrm{max}-y) -
                  \Lambda_\mathrm{min}}\,\rmd y \ ,
\end{equation}
where the lower integration bound is
\begin{equation}
y_\mathrm{inf}=\max\left\lbrace \Lambda_\mathrm{min},\,\frac{x-\gamma
    \Lambda_\mathrm{max}}{1-\gamma}\right\rbrace \ ,
\end{equation}
representing the smallest $y$ from where $x$ can be reached in a
single jump; the second term in brackets is obtained by inverting the
expression of the soft bound, Eq.~\eqref{eq:additive_softbound},
$x=y+\gamma\left(\Lambda_\mathrm{max}-y\right)$.  Outside the interval
$[\Lambda_\mathrm{min},\,\Lambda_\mathrm{max}]$, $\rho(x)$ is
identically zero.  In order to simplify the formulas, and without loss
of generality, we can fix $\Lambda_\mathrm{min} = 0$ and
$\Lambda_\mathrm{max} = 1$ (we will consistently use this convention
in the remainder of this section).  Hence, $\rho(x)$ becomes
\begin{equation}\label{eq:def4}
	\rho(x) = \int_{\max\left\lbrace 0,\,\frac{x-\gamma}{1-\gamma}\right\rbrace}^{1} 
		\frac{\rho(y)}{y+\gamma(1-y)}\,\rmd y,
\end{equation}
or equivalently
\begin{equation}\label{eq:def4b}
	\rho(x) = \rho_0 - \int_0^{\max\left\lbrace 0,\,\frac{x-\gamma}{1-\gamma}\right\rbrace}
			\frac{\rho(y)}{y+\gamma(1-y)}\,\rmd y \ ,
\end{equation}
where $\rho_0 \equiv \int_{0}^{1} \frac{\rho(y)}{y+\gamma(1-y)}\,\rmd
y$.  Note that in this case $\rho(x)$ depends on $x$ only through the
boundaries of the definite integral in Eq.~(\ref{eq:def4b}).  Therefore,
since the integral depends only on $0 < y < \frac{x-\gamma}{1-\gamma}$
for any given $x$, Eq.~(\ref{eq:def4b}) translates into an iterative
procedure for computing the stationary distribution $\rho(x)$,
yielding the piecewise analytical solution for adjacent intervals.  In
fact, for $x < \gamma$ (let us call it interval I), the density
distribution is constant and its value is $\rho(x) = \rho_0$, because
the upper integration bound in Eq.~(\ref{eq:def4b}) is zero.  Now,
interval II is defined as the region for which $\rho(x)$ can be
calculated from (\ref{eq:def4b}) solely in terms of $\rho\left(y\in
  \mathrm{interval}\;\mathrm{I}\right)$, namely $\gamma < x <
1-(1-\gamma)^2$.  Iterating this procedure, the $n$-th interval is
found to be $1-(1-\gamma)^{n-1}<x<1-(1-\gamma)^n$, and the analytical
form of $\rho(x)$ inside it can be calculated in terms of the $n-1$
solutions already obtained.  A few steps are explicitly presented in
the Appendix.  Figure \ref{figure:exact_BOX} shows the perfect
accordance of the analytical solution with Monte Carlo simulation.

\begin{figure}
\centering
\includegraphics[scale=0.22]{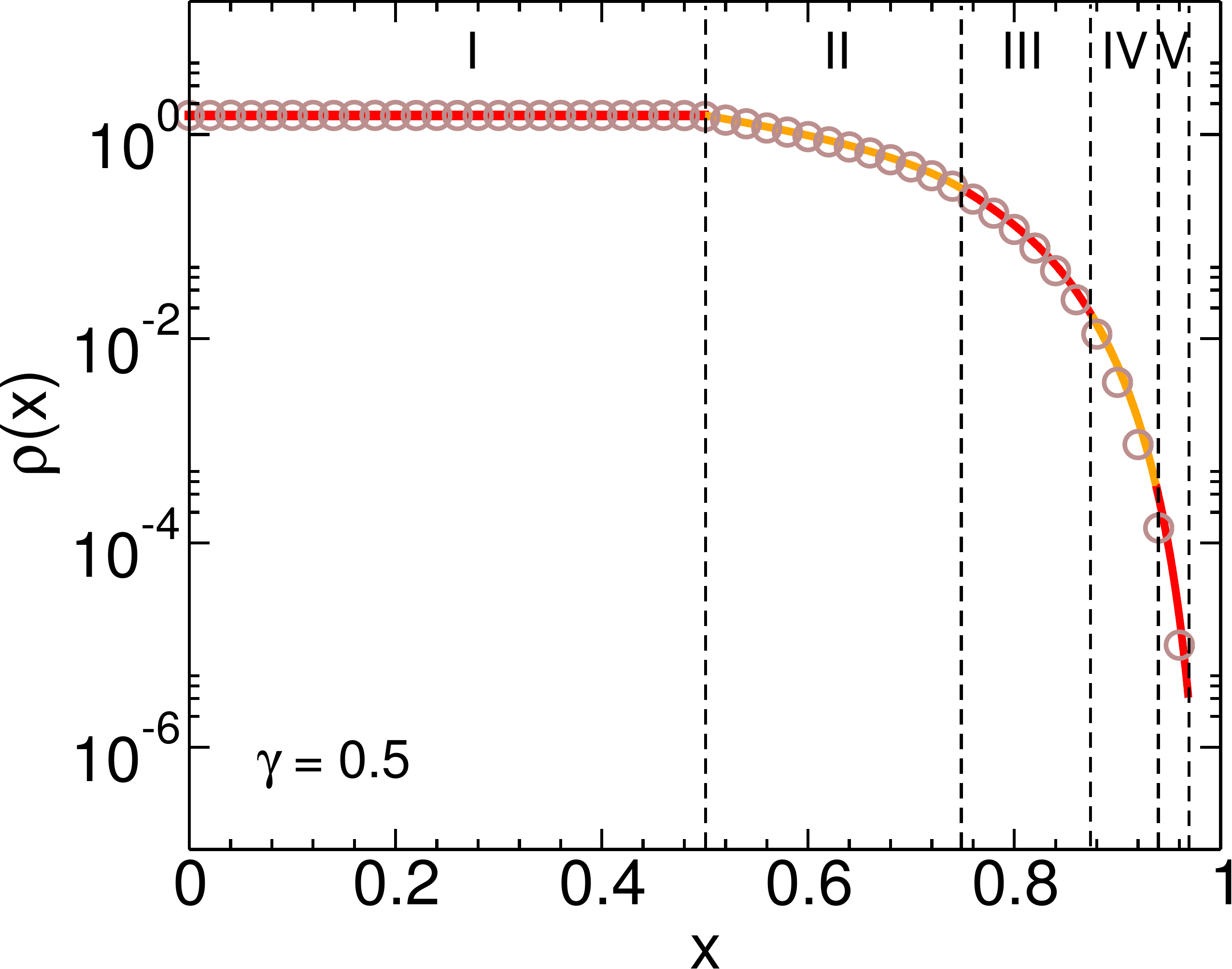}
\caption{\label{figure:exact_BOX} (Color online) The stationary state $\rho(x)$
    can be computed analytically by iterative recursion in the case of
    an underlying flat transition probability. The $n$-th step in
  the calculation gives the analytical steady state (solid lines) in
  all intervals up to the $n$-th.  (Operatively, one must resort to
  numerical integration after interval III, see the Appendix.)
  Circles indicate the results from numerical Monte Carlo simulations
  (with $10^6$ particles).  Notice the logarithmic scale on the
  vertical axis.  }
\end{figure}

\begin{figure}
\centering
\includegraphics[scale=0.22]{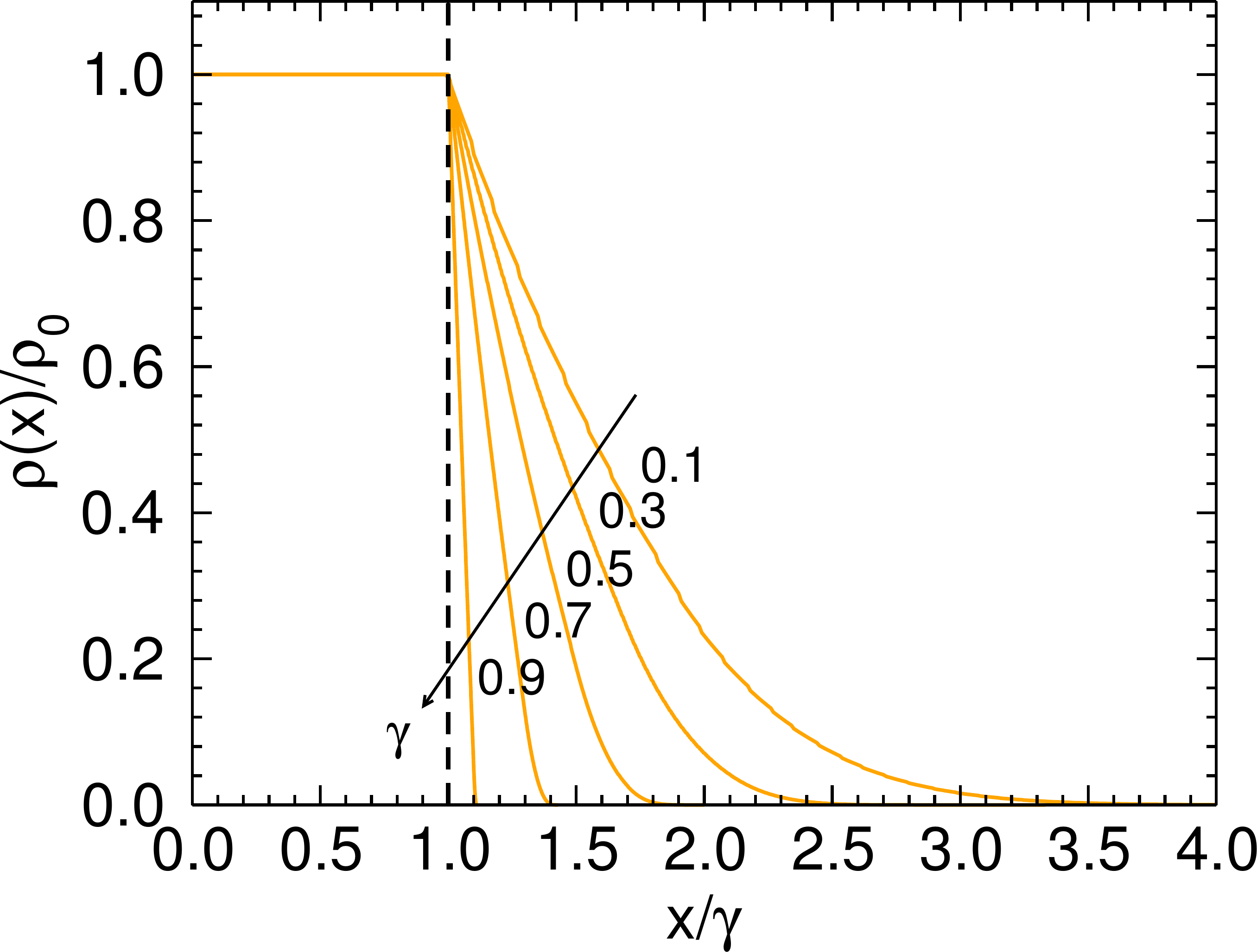}
\caption{\label{figure:comparison_BOX} (Color online) The stationary state is
    flat in the region $(0,\gamma)$.  The shoulder shrinks by
    increasing the parameter $\gamma$. The curves are the steady
  state distributions (computed by numerical integration) for the case
  of flat transition probability and soft bounds, plotted as a
  function of $x/\gamma$, for several values of $\gamma$.  The
  probability density $\rho$ is rescaled by its value in the origin,
  in order to better compare the different curves. As expected, when
  the upper bound becomes a hard bound ($\gamma\to 1$), one recovers
  the usual constant solution for reflective boundary conditions
  \cite{Risken:BOOK}.  }
\end{figure}

It is important to stress here that the behavior of the steady state
for a soft bound is qualitatively distinct from a hard bound.  As
noted before, the stationary state is constant for $0<x<\gamma$
(interval I) and then decreases quickly towards zero in the following
intervals.  Viewed in logarithmic scale, its shape displays a
characteristic shoulder starting at $x=\gamma$, which would be absent
if the upper bound were hard (in which case $\gamma=1$).  A
qualitative comparison of the slope of the shoulder as a function of
$\gamma$ can be obtained if one collapses different plots, at
different values of $\gamma$, by rescaling $x\mapsto x/\gamma$ and
$\rho(x)\mapsto \rho(x)/\rho_0$.  The result is shown in linear scale
in Fig.~\ref{figure:comparison_BOX}.

\paragraph{Entropy production.}\label{par:entr_production}
As observed above, the diffusion within a soft bound is irreversible;
this is signaled by a non-zero entropy production.  The
entropy production per time step at stationarity~\cite{Derrida:2007}
can be defined for the transition between two states $x$ and $y$ as
\begin{equation}
  \Delta S_{x,y} = k \ln \frac{\rho(x)\mathcal{P}_{x\to y}}{\rho(y)\mathcal{P}_{y\to x}},
\end{equation}
where $k$ is Boltzmann's constant.  It can be evaluated easily for two
states belonging to interval I in Fig.~(\ref{figure:exact_BOX}), where
$\rho(x)/\rho(y)=1$; the difference between the transition
probabilities comes from the normalization factor
\begin{equation}
\label{eq:entropy}
\Delta S_{x,y\in\mathrm{I}} = k \ln \frac{y+\gamma(1-y)}{x+\gamma(1-x)}.
\end{equation}
Since this expression is different from zero in general,
  Eq.~(\ref{eq:entropy}) shows that the Markov chain is irreversible
in a regime where neither $\mathcal{P}_{x\to y}$ nor
$\mathcal{P}_{y\to x}$ are vanishing, thus complementing the argument
given at the beginning of this Section.  For small jumps
$y=x+\epsilon$
, the entropy (\ref{eq:entropy}) at first order in $\epsilon$ results
\begin{equation}
  \Delta S_{x,x+\epsilon}\approx k \epsilon \left(x+\frac{\gamma}{1-\gamma}\right)^{-1},
\end{equation}
which is always positive.  Consequently, there is a local net
imbalance towards larger sizes, which is due solely to the different
volumes of configuration space available to different states; in fact,
the bulk of the transition probability, close to $x=y$, is symmetric.
This tendency is largest for $x$ close to the lower bound, and
decreases for larger values of the variable.  We point out that a
similar trend in the evolution of mammalian body masses (called
\emph{Cope's law} in this context) has been long studied and debated
\cite{Cope:1887,Gould:1997,Clauset2008}, although it is usually
ascribed to an asymmetry or a drift in the bulk transition
probability.  In the soft-bound framework, this feature has entropic
origins and emerges naturally.

\begin{figure}[t]
\centering
\includegraphics[scale=0.22]{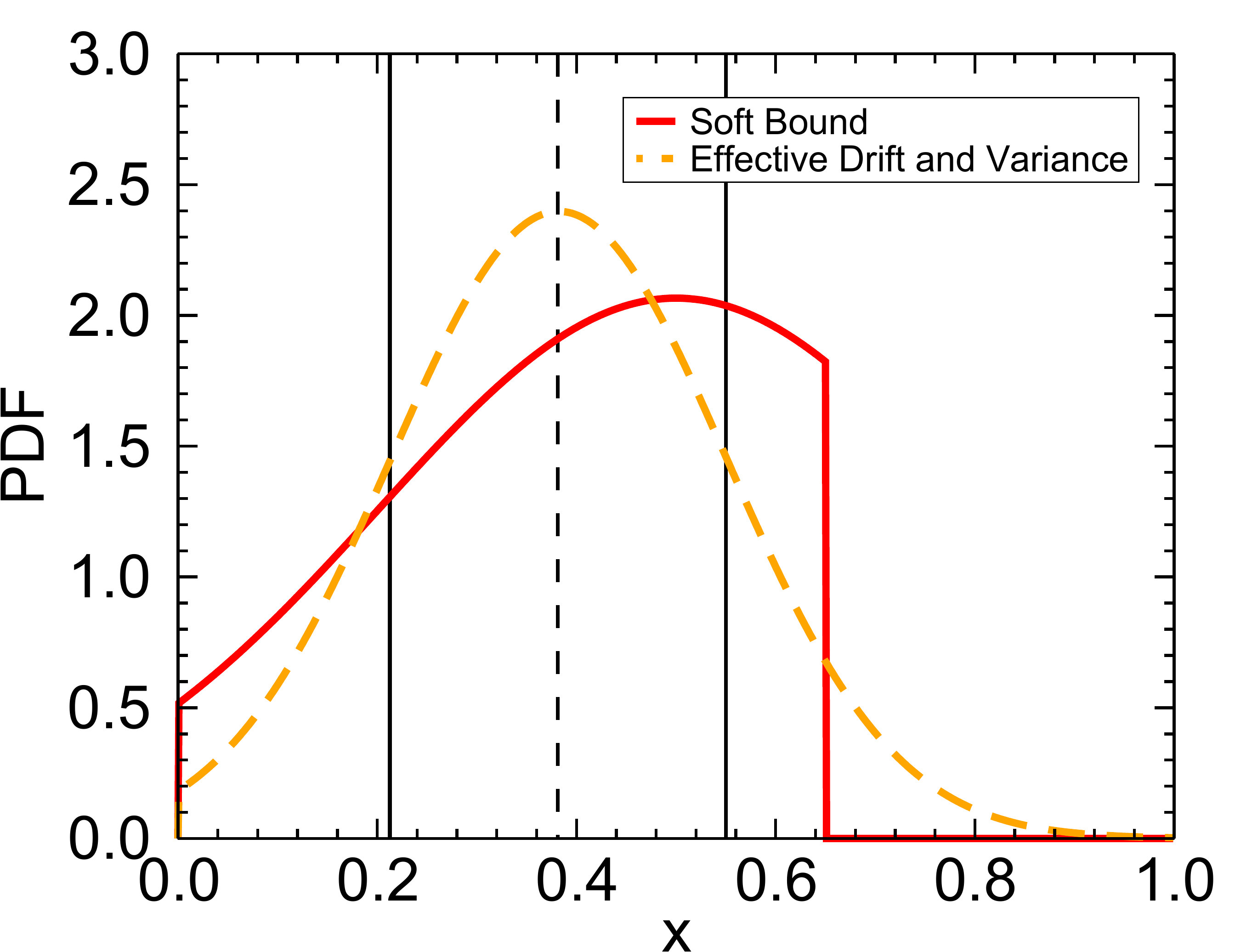}
\caption{\label{figure:example_eff} (Color online) Approximation of the soft
    bound by an effective Gaussian distribution
    with hard bounds. The Gaussian jump distribution
  $\mathcal{P'}_{y\to x}$ (dashed curve) has mean $\tilde\mu$
  (vertical dashed line) and variance
  $\tilde\sigma^2$ (represented by the vertical solid lines) computed
  from $\mathcal{P}_{y\to x}$ (solid curve), and is bounded in a
  size-independent way.}
\end{figure}

\paragraph{The stationary distribution is not simply the consequence
  of effective drift and variance.}
We now explore in further depth the qualitative differences between a
hard and a soft bound.  The simple case of a flat $\pi(x-y)$
illustrates the fact that the transition kernel $\mathcal{P}_{y\to x}$
for a given position $y$ is asymmetric in $x$, and the asymmetry
depends on $y$.  It is then natural to ask whether the same form of
the steady state could in principle be obtained by using a symmetric
jump distribution $\mathcal{P}^\prime_{y\to x}$ with only hard bounds,
but adding an effective position-dependent drift and a variance.

To illustrate this point, we consider the situation represented in
Fig.~\ref{figure:example_eff}.  Here, the jump distribution
$\mathcal{P}_{y\to x}$ (solid curve) is built as in
Eq.~(\ref{eq:kernel_product}), starting from a Gaussian underlying
probability $\pi(x-y)$ with mean zero and variance $\sigma^2$.  As a
consequence of the bounding kernel $\beta(x,y)$, the mean and variance
of $\mathcal{P}_{y\to x}$ will be different, and they will have a
dependence on $y$ (and of course on the softness parameter $\gamma$).
Let us call them $\tilde\mu(y,\gamma)$ and $\tilde\sigma^2(y,\gamma)$
respectively; they are defined as
\begin{equation}\label{eq:mu_eff}
	\tilde \mu(y,\,\gamma) = \int_0^1 \mathcal{P}_{y\to x}\,x \,\rmd x
\end{equation}
and
\begin{equation}\label{eq:sigma_eff}
	\tilde\sigma^2(y,\,\gamma) = \int_0^1\,\mathcal{P}_{y\to x}\,x^2\,\rmd x - \tilde\mu^2(y,\gamma).
\end{equation}
In Fig.~\ref{figure:example_eff}, mean and variance are represented
respectively by the vertical dashed line and the two vertical solid
lines.  Note that in general $\tilde\mu(y,\,\gamma)\not=0$ and
$\tilde\sigma^2(y,\,\gamma) \not= \sigma^2$.  The effective
distribution $\mathcal{P'}_{y\to x}$ (dashed curve in
Fig.~\ref{figure:example_eff}) is constructed as a Gaussian with mean
$\tilde\mu$, variance $\tilde\sigma^2$ and lower and upper hard
bounds:
\begin{equation}\label{eq:distr_jump_eff}
\mathcal{P}^\prime_{y\to x} = \mathcal{Z}^{-1}\chi_{[0,\,1]}(x)\,
\exp\left[- \frac{\left(y - \tilde\mu\right)^2}{2 \tilde\sigma^2}\right].
\end{equation}
Note that $\mathcal{P}^\prime_{y\to x} \not=
\mathcal{P}_{y\to x}$ and, more importantly, $\mathcal{P}^\prime_{y\to x}$ 
does not have any soft bound. 

\begin{figure*}
\centering
\includegraphics[scale=0.2]{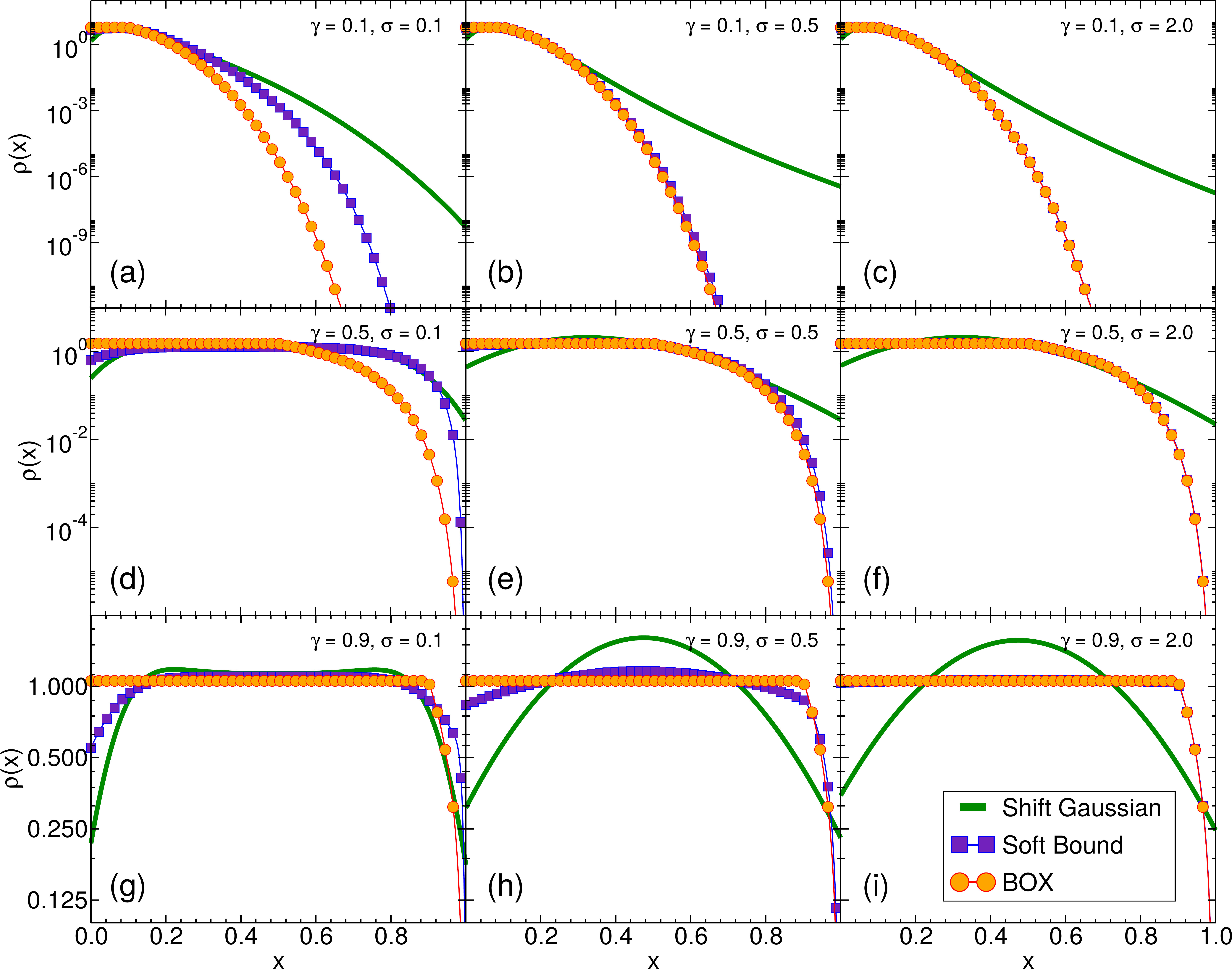}
\caption{\label{figure:comparison_effective} (Color online) The steady state
    produced by the soft bound (blue squares) is not realized by a
    discrete Gaussian diffusion with effective drift and variance
    (green lines). Large values of $\sigma$ give similar results to
  the analytical solution (orange disks) for a flat jump distribution.
  Lower values of $\gamma$ give rise to increasingly skewed
  distributions.  }
\end{figure*}

In order to understand if the effects of a soft bound can be recovered
by using effective drift and variance, we
numerically studied the steady state using both $\mathcal{P}_{y\to
  x}$ and $\mathcal{P}^\prime_{y\to x}$ by varying the variance
$\sigma^2$ and the soft bound $\gamma$, 
also in the limit when
the soft bound becomes a hard bound.
Figure~\ref{figure:comparison_effective} shows that the steady state of
the effective jump distribution $\mathcal{P}^\prime_{y\to x}$ (green
lines) is very different, for most 
choices of $\sigma$ and $\gamma$,
from that obtained using the true soft bound $\mathcal{P}_{y\to x}$
(blue squares). Hence, the effective drift model is in general
ineffective. It becomes an acceptable approximation only
when $\sigma^2$ is small and $\gamma$ is large.
In particular, the shoulder characteristic of the soft bound is never
well reproduced by the effective process with hard bound and drift.

Figure~\ref{figure:comparison_effective} also shows the conditions for
which the (analytically computable) steady-state distribution for a
soft-bound process with flat transition probability (orange disks in
figure) is a good approximation of the soft-bound jump process with a
Gaussian transition probability.
In brief, for fixed $\gamma<1$, the true steady state approaches (i)
the results of the effective drift model when $\sigma\to 0$, and (ii)
those for the box distribution when $\sigma\to\infty$.  This behavior
is easily rationalized by the two facts that (i) the soft bound affects
the tail of the jump distribution, and (ii) the Gaussian kernel becomes
effectively flat for large variance.  The accord for intermediate
values of $\sigma$ depends on $\gamma$, and improves for larger values
of this parameter (i.e., for harder bounds).

\section{Dynamics of the maximum and Gompertz law}
\vspace{-0.3cm}

The nature of the soft bound also affects the relaxation to the steady
state.  Given an initial distribution $\rho_0(x)$, supported in a
subinterval of
$\left[\Lambda_\mathrm{min},\Lambda_\mathrm{max}\right)$, one can
study the time dependence of the maximum $X_t$, i.e., of the largest
$x$ on which $\rho_t(x)$ is non zero.  This may serve as an easily
accessible empirical observable, which often turns out to be relevant
to characterize the system.  For instance, in the evolution of
software package sizes, the rightmost point in the support of
$\rho_t(x)$ represents the evolution of (the logarithm of) the ``largest package size'' in the
operating system. For mammalian body masses, this quantity represents
the logarithm of the mass $M_t$ of the largest species at a given time
$t$, which is the object of much attention in paleobiology, as it can
hold information on macro-evolutionary patterns~\cite{Smith:2010}.  For
mammals it has been observed, perhaps surprisingly, that this maximum
is not dominated by a single taxon nor by a single continent. 
Different ecological and evolutionary approaches have been applied to
the evolution of the maximum mammalian mass.  In particular, an
unconstrained multiplicative diffusion predicts \cite{Trammer:2005}
that it grows indefinitely as $X_t=\log M_t\sim t^{1/2}$.
Another model is based on the Gompertz law, a particular logistic
function originally proposed as a phenomenological description of
mortality in a population \cite{Gompertz:1825} (used also in the
context of tumor growth \cite{Speer:1984,AlbanoGiorno:2006}), and
tries to capture in an empirical way the overall effects of
constraints, and assumes the following saturating evolution (we use
the notation of ref.~\cite{Smith:2010}):
\begin{equation}
\label{eq:gompertz}
\log M=\log K - \log\frac{K}{M_0} e^{-\alpha t},
\end{equation}
where $M_0$ is the maximum mass at time 0, $K$ plays the role of a
carrying capacity, and $\alpha$ is a
characteristic exponent.  We show in this section that a Gompertz law
for the maximum emerges as a natural consequence of the soft-bound
mechanism.

Since we are interested in the evolution of the maximum, the actual
shape of $\pi(x-y)$ is not important, as long as its support contains
that of $\beta(x,y)$, so that the soft upper bound is the only
responsible for the dynamics of $X$.  Let $X_0=\log M_0$ be the
maximum at time $0$.  Because $\mathcal{P}_{y\to x}$ is bounded 
by Eq.~(\ref{eq:additive_softbound}), the maximum $X_1$ that can be
reach at time $t=1$ will be
\begin{equation}
	X_1 = \Lambda_\mathrm{max} + (1-\gamma)(X_0 - \Lambda_\mathrm{max}).
\end{equation}
At time $t = 2$, the maximum position that can be reached is
\begin{equation}
\begin{aligned}
	X_2 &=\Lambda_\mathrm{max} + (1-\gamma)(X_1 - \Lambda_\mathrm{max})\nonumber\\
            &= \Lambda_\mathrm{max} + (1-\gamma)^2 (X_0 - \Lambda_\mathrm{max}).
\end{aligned}
\end{equation}
Therefore, the maximum position at time $t$ will be
\begin{equation}\label{eq:log_gompertz}
	X_t = \Lambda_\mathrm{max} + (1-\gamma)^t (X_0 - \Lambda_\mathrm{max}).
\end{equation}
This equation is the logarithmic version of the Gompertz law, i.e.,
the corresponding law for additive diffusion.  Indeed, writing
Eq.~(\ref{eq:log_gompertz}) explicitly for $M_t$ yields
\begin{equation}
\log M_t = \Lambda_\mathrm{max} - \log\frac{M_0}{K} e^{\log(1-\gamma)t},
\end{equation}
which is the Gompertz law Eq.~(\ref{eq:gompertz}) with
$K=\exp(\Lambda_\mathrm{max})$ and $\alpha=-\log(1-\gamma)$.  We
stress that this result is valid independently of the underlying
jump distribution $\pi(x-y)$:
it is simply the consequence of the
functional form of the soft bound, and thus it is very general.
Quantitatively, the value $\gamma\approx 0.2$ measured from a
compilation of $1109$ ancestor-descendant mass ratios for North
American terrestrial mammals \cite{Alroy1998,Gherardi:2013PNAS} yields
an estimate $\alpha\approx 0.2$.  This is in line with the results
($\alpha\approx 0.1$) obtained from fossil data on the evolution of
the largest mammalian mass \cite{Smith:2010}.

\begin{figure}[t]
\centering
\includegraphics[scale=0.22]{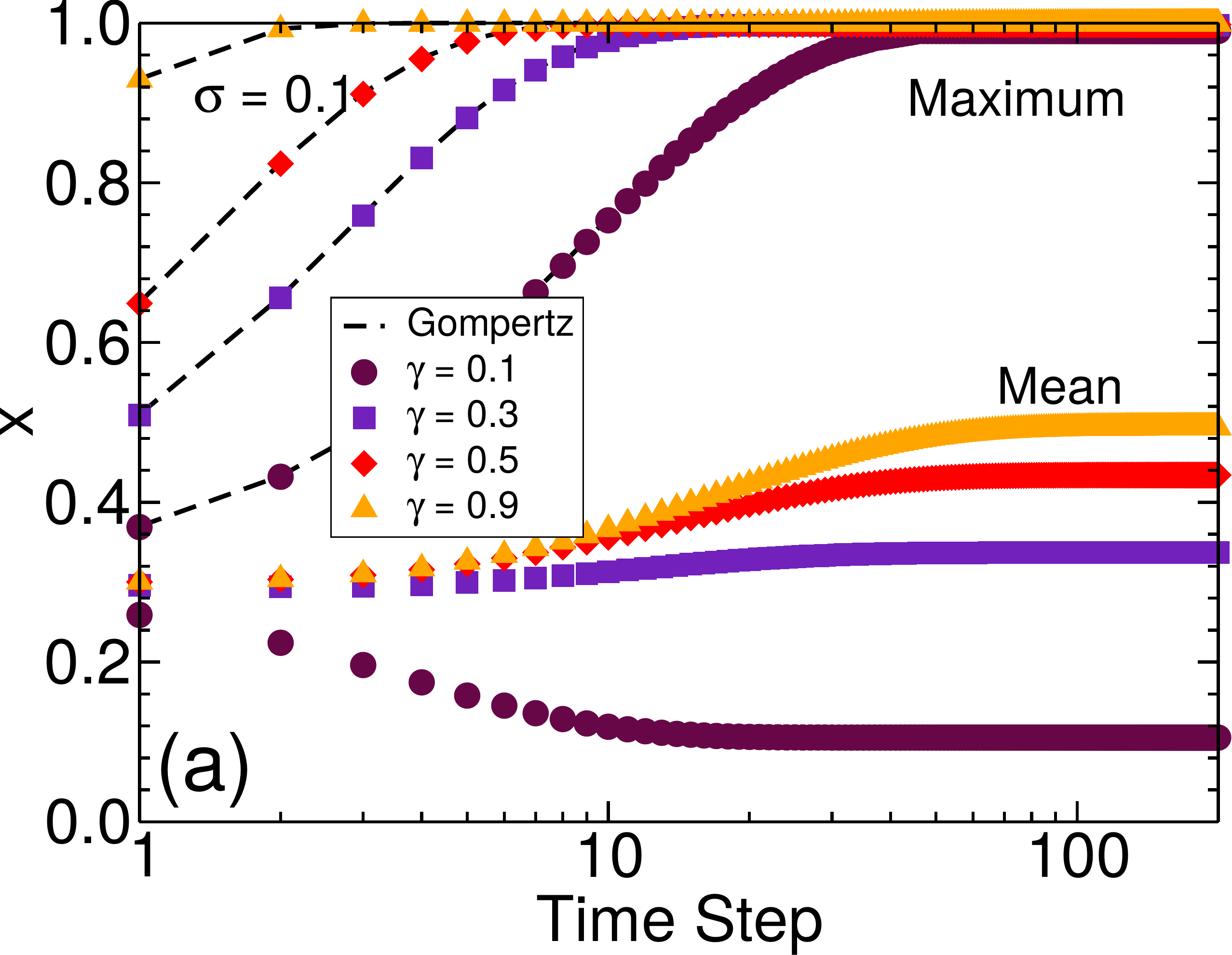}
\includegraphics[scale=0.22]{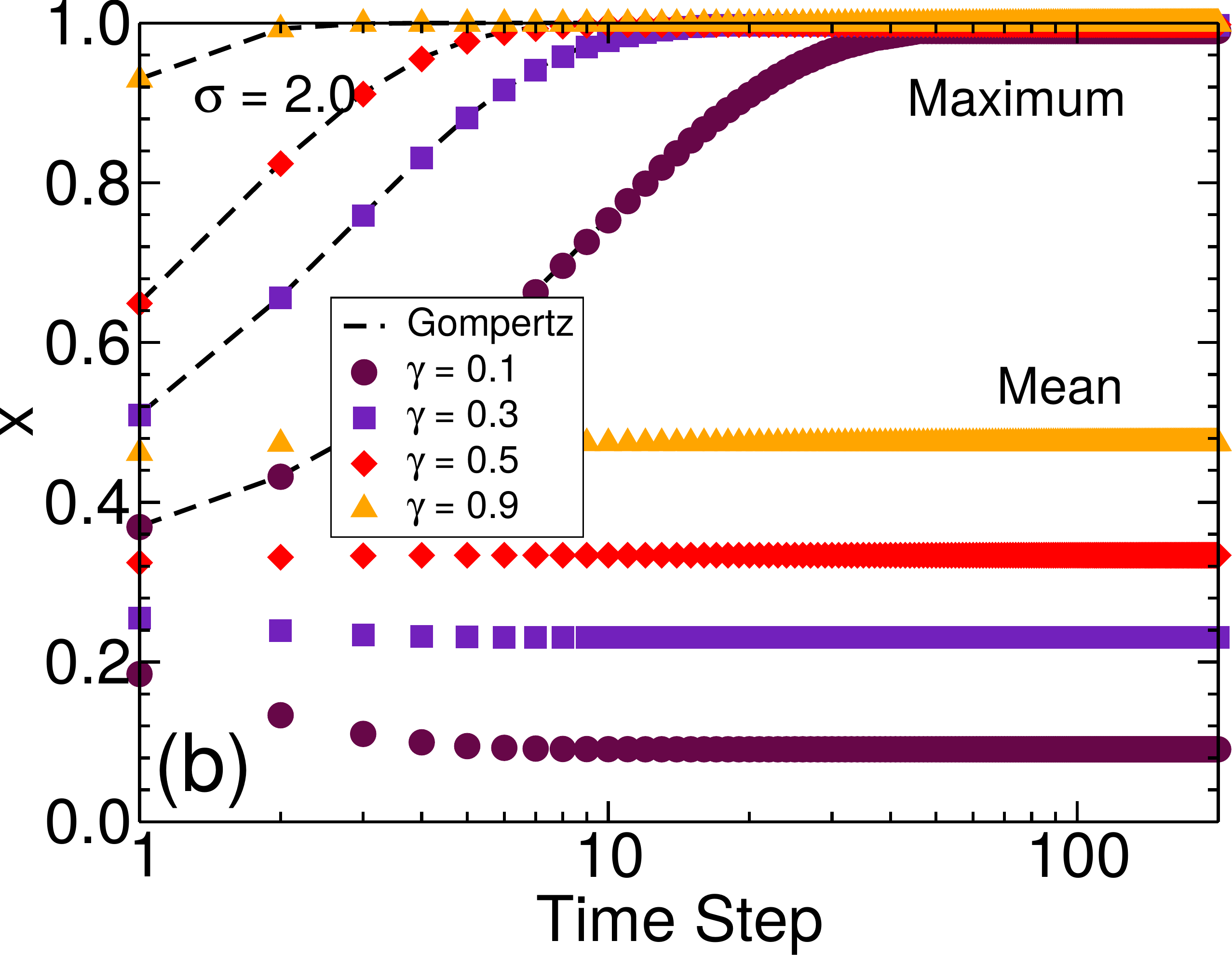}
\caption{\label{figure:gompertz} (Color online) The dynamics of the maximum
    follows precisely the Gompertz function, as is shown analytically
    in the text.
The dynamics of the maximum is independent of the shape of the jump
distribution, 
and only depends on $\gamma$.
On the contrary, the dynamics and the stationary value
of the mean depend on the details of the distribution;
$\sigma=0.1$ in panel (a), $\sigma=2$ in panel (b).
Dashed lines are Gompertz functions, symbols are simulations.
}
\end{figure}

Figure~\ref{figure:gompertz} compares the analytical solution of
Eq.~(\ref{eq:log_gompertz}) 
with numerical simulations.  The accordance
between the two is remarkable. Note that a finite population of $N$
particles does not in general attain exactly the predicted maximum,
but finite-size deviations are expected for small $N$.  In practice,
we expect these errors to likely be negligible for both Ubuntu
packages ($N\approx 40\,000$) and mammalian body masses ($N\approx
4000$ in the MOM dataset~\cite{MOM}, used
in Ref.~\cite{Gherardi:2013PNAS}).  Also note that, contrary to the
maximum, the mean size can either increase or decrease with time in
this process, depending on the parameters and on the initial
conditions.

\section{Discussion and Conclusions}
\vspace{-0.3cm}

Having shown that the soft bound mechanism is qualitatively distinct
from hard bounds, 
it will be important to
determine whether other empirical systems show
features that are compatible with the existence of soft bounds.
We have analyzed here two main signatures of a soft
bound. The first is the formation of a non-trivial shoulder in the
steady-state distribution, and the second is the Gompertzian growth of
the maximum.  These two signatures can be used in practical
applications as ``smoking guns'' for this kind of
behavior. Importantly, the soft bound mechanism can be relevant only
when the underlying diffusion process has intrinsically discrete
nature.  We speculate that the soft bound mechanism can occur in
situations where the concerted action of many degrees of freedom is
proxied by low- or one-dimensional variables (such as ``size'' or
``mass'').  In this view, a single jump in size can be seen as the
result of a large number of changes in a high-dimensional parameter
space, each subject to complex hard bounds (which are more natural to
picture), which concur to give rise to the soft bound phenomenon.

Another remarkable feature of a discrete-time diffusion process with
soft bounds is the non-reversibility (somewhat analogous to the
asymmetry found in the kinetic proofreading \cite{BarZiv:2002}),
giving rise to a probability flux of entropic nature. Interestingly,
this entropic unbalance, which is naturally present in our model, can
provide an alternative (purely entropic) explanation of the Cope's
rule for mammalian evolution.

To conclude, we address a possible generic mechanism
that could give rise to soft bounds.  Let us consider a complex
interacting system with many components or agents, where a scalar
order parameter $s$, which we can term ``size,'' effectively follows a
discrete diffusion process.  This variable can be for instance the
number of lines of a software project, or the mean mass of an animal
species, the number of workers in a firm or the number of inhabitants
in a city.  We suppose that there exists a function, similar to a
power in non-equilibrium thermodynamics, estimating the ``effort'' $E$
that is put into the system for a given span of time. For example the
effort can be proxied by the total man-hours spent on the code by
programmers, or the food intake of an animal, or the money spent by a
firm or city administration.  We further assume that a scaling
relation
\begin{equation}
\label{eq:allometric_E}
E\propto s^\alpha
\end{equation}
holds between effort and size, where $\alpha$ is an exponent
associated to $E$.  This assumption can be seen as an instance of
\emph{allometric scaling}, a feature commonly found in complex systems,
where some quantity has a power-law dependence on size.
It is observed for instance in general ontogenetic growth and the
metabolic rates of animals \cite{Brody:BOOK,vonBartalanffy:1957}
(albeit with some deviations \cite{Kolokotrones:2010}), transportation
networks \cite{BanavarMaritan:1999}, and city organization
\cite{MarsiliZhang:1998,Bettencourt:2007,LoufBarthelemy:2013}.
This relation expresses the principle that the total amount of effort
available for the system per unit time can scale sub- or
super-linearly with size, with respectively $\alpha < 1$ and $\alpha > 1$; 
for the case at hand we suppose $\alpha<1$.
Note that Eq.~(\ref{eq:allometric_E}), in a thermodynamic interpretation, is
different from a Green-Kubo relation where $\alpha=1$ strictly.

We assume the underlying hypothesis that the effort flow is used both
for maintaining the system and for increasing its size; the maximum
increase, corresponding to exhausting all available effort, thus
satisfies
\begin{equation}
s^\alpha = c_\mathrm{m} s + c_\mathrm{d} \left( s' - s\right),
\end{equation}
where the two constants $c_\mathrm{m}$ and $c_\mathrm{d}$ are the
efforts per unit size needed for maintenance and development
respectively.  Therefore the maximum multiplicative size change
attainable in a given time span is
\begin{equation}
  \label{eq:allometric_bound}
  \frac{s'}{s}=
  \frac{s^{\alpha-1}}{c_\mathrm{d}}
  +\left(1-\frac{c_\mathrm{m}}{c_\mathrm{d}}\right). 
\end{equation}
If the two cost constants are similar ($c_\mathrm{d}\approx
c_\mathrm{m}$), or when $s'/s$ is large,
Eq.~(\ref{eq:allometric_bound}) is just the soft bound
Eq.~(\ref{eq:multiplicative_softbound}), with $\gamma=1-\alpha$ and
$s_\mathrm{max}=\left(c_\mathrm{d}\right)^{-1/\gamma}$. In brief, this
argument shows that a scaling hypothesis on the relation between
effort-rate and size implies a multiplicative soft bound (similar to
the one suggested by both software and mammal mass).  We anticipate
that the ``effort'' is not necessarily an abstract quantity, but can
be possibly measured in different systems. For example, data is
available on the number, extent and frequency of updates of software
packages. Thus, the above argument may be (in line of principle)
testable, and opens a question for future work.

\begin{acknowledgments}
\vspace{-0.3cm}
 
We acknowledge useful discussions with Bruno Bassetti
and Amos Maritan.
M.G.\ was partially supported by Fondo Sociale Europeo
(Regione Lombardia) through the grant ``Dote Ricerca.'' S.M.\ 
acknowledges the Air Force Office of Scientific Research (prime sponsor) 
and the University of California, San Diego, for the partial support 
to this work under the grants FA9550-12-1-0046 and 10323836-SUB.
\end{acknowledgments}

\onecolumngrid
\appendix*
\section{}

This appendix details some steps of the iterative procedure presented
in Sec.~\ref{section:flat_distribution}, for computing the stationary
state in the case of a flat jump probability distribution.  Starting
from Eq.~(\ref{eq:def4b}), the solution in the interval $0<x<\gamma$
is the constant $\rho(x)=\rho_0$.  For $\gamma < x < 1-(1-\gamma)^2$
(interval II) one obtains
\begin{align}\label{eq:step2}
	\rho(x) &= \rho_0 - \int_0^{\frac{x-\gamma}{1-\gamma}}\frac{\rho(y)}{y+\gamma(1-y)}\,\rmd y\nonumber\\
			&= \rho_0\left[1 - \int_0^{\frac{x-\gamma}{1-\gamma}}\frac{1}{y+\gamma(1-y)}\,\rmd y\right]\nonumber\\
			&= \rho_0\left[1-\frac{\log(x/\gamma)}{1-\gamma}\right]
			= \rho_0\,\Bigg[1-\rho_1(x)\Bigg].
\end{align}
The function $\rho_1(x)=\log(x/\gamma)/(1-\gamma)$
corresponds to the distribution at time $t=1$ obtained by starting from a constant
initial distribution supported on interval I.

Interval III is defined by $1-(1-\gamma)^2 < x < 1-(1-\gamma)^3$, and in this region
$\rho(x)$ can be calculated as
\begin{align}
\rho(x) &=
\rho_0 \left[1 - \int_0^{\gamma}\frac{1}{y+\gamma(1-y)}\,\rmd y
 -  \int_\gamma^{\frac{x-\gamma}{1-\gamma}}\frac{1-\rho_1(y)}{y+\gamma(1-y)}\,\rmd y\right]\nonumber\\
&= \rho_0\left[1 - \int_0^{\frac{x-\gamma}{1-\gamma}}\frac{1}{y+\gamma(1-y)}\,\rmd y 
 +  \int_\gamma^{\frac{x-\gamma}{1-\gamma}}\frac{\rho_1(y)}{y+\gamma(1-y)}\,\rmd y\right]
= \rho_0\,\Bigg[1-\rho_1(x) - \rho_2(x)\Bigg],
\end{align}
where
\begin{align}\label{eq:app_rho2}
	\rho_2(x) &= -\int_\gamma^{\frac{x-\gamma}{1-\gamma}}\frac{\rho_1(y)}{y+\gamma(1-y)}\,dy
              = \frac{\log\left(\frac{x}{\gamma}\right)\log\left(\frac{\gamma(1-\gamma)}{x-\gamma}\right)}
              		{(1-\gamma)^2}
               +\frac{\text{Li}_2(\gamma-1)-\text{Li}_2(1-\frac{x}{\gamma})}{(1-\gamma)^2},
\end{align}
which corresponds to the solution at time $t=2$. Unfortunately, due to the presence of dilogarithm functions
$\text{Li}_2(\cdot)$ in Eq.~(\ref{eq:app_rho2}), the full analytical tractability of $\rho(x)$ is broken for
further intervals.

\twocolumngrid

\end{document}